\documentclass[letterpaper, 10 pt, conference]{ieeeconf}

\IEEEoverridecommandlockouts

\usepackage[utf8]{inputenc}
\usepackage[T1]{fontenc}
\usepackage{microtype,newpxtext,newpxmath}
\usepackage{mathtools,gensymb,siunitx,eurosym,url,soul}
\usepackage[font=footnotesize,textfont=footnotesize]{subcaption}
\usepackage[font=small,textfont=small]{caption}
\usepackage{longtable,array,booktabs,mdwlist}
\usepackage{comment,todonotes,lineno,acronym}
\usepackage{xcolor,color}
\usepackage{graphicx}
\usepackage{lineno}
\usepackage{algorithmic,mwe,units}
\usepackage{hyperref}

\hypersetup{colorlinks=true, breaklinks=true, pagebackref=true,
  urlcolor=blue, linkcolor=blue,anchorcolor=blue,citecolor=blue,
  pdfpagemode = UseNone, 
  pdfauthor = {},
  pdftitle = {},   
  pdfsubject = {},
  pdfkeywords = {}
}

\setkeys{Gin}{draft=false} 
\graphicspath{{img/} {./}}
\acrodef{EMG}[EMG]{electromyography}
\acrodef{EEG}[EEG]{electroencephalography}
\acrodef{ECG}[ECG]{electrocardiography}
\acrodef{HFO}[HFO]{High-Frequency Oscillations}
\acrodef{LFP}[LFP]{Local Field Potential}
\acrodef{BMI}[BMI]{Brain-Machine Interface}
\acrodef{SNN}[SNN]{Spiking Neural Network}
\acrodef{RNN}[RNN]{Recurrent Neural Network}
\acrodef{SRNN}[SRNN]{Spiking Recurrent Neural Network}
\acrodef{CNN}[CNN]{Convolutional Neural Network}
\acrodef{AER}[AER]{Address Event Representation}
\acrodef{ANN}[ANN]{Artificial Neural Network}
\acrodef{FPGA}[FPGA]{Field Programmable Gate Array}
\acrodef{DPI}[DPI]{Differential Pair Integrator}
\acrodef{CPG}[CPG]{Central pattern generator}
\acrodef{rCPG}[rCPG]{respiratory CPG}
\acrodef{IF}[IF]{Integrate-and-Fire}

\title{\LARGE \bf
  Neuromorphic Pattern Generation Circuits for Bioelectronic Medicine
}

\author{Elisa Donati$^{1}$, 
Renate Krause$^{1}$, and 
Giacomo Indiveri$^{1}$
\thanks{*This work was supported by the EU-H2020 FET project CResPACE (Grant No. 732170)}
\thanks{$^{1}$Institute of Neuroinformatics, University of Zurich and ETH Zurich, Zurich, Switzerland \small [elisa$|$rekrau$|$giacomo]@ini.uzh.ch}}%

\begin{document}
\maketitle
\thispagestyle{empty}
\pagestyle{empty}

\begin{abstract}
  Chronic diseases can greatly benefit from bioelectronic medicine approaches.
  Neuromorphic electronic circuits present ideal characteristics for the development of brain-inspired low-power implantable processing systems that can be interfaced with biological systems.
  These circuits, therefore, represent a promising additional tool in the tool-set of bioelectronic medicine.
  In this paper, we describe the main features of neuromorphic circuits that are ideally suited for continuously monitoring the physiological parameters of the body and interact with them in real-time.
  We propose examples of computational primitives that can be used for real-time pattern generation and present a neuromorphic implementation of neural oscillators for the generation of sequence activation patterns.
  We demonstrate the features of such systems with an implementation of a three-phase network that models the dynamics of the respiratory Central Pattern Generator (CPG) and the heart chambers rhythm, and that could be used to build an adaptive pacemaker.
\end{abstract}

\section{INTRODUCTION}

Chronic diseases are one of the leading causes of death in the world population. The majority of deaths caused by these diseases are attributed to respiratory failure, cardiovascular disease, cancers, and diabetes~\cite{Sav_etal15}.
As the proper treatment and management of chronic conditions is still an open challenge, there is a growing interest in implantable electronic devices that can monitor physiological parameters and control these in a closed-loop interaction with the body in real-time. 
In addition, the very same technology developed for these implantable devices used to support the treatment of chronic diseases can be used for controlling closed-loop prosthetic devices and brain-machine interfaces.

A promising approach toward the development of this technology is the one of ``neuromorphic engineering''~\cite{Mead20,Indiveri_Horiuchi11}.
Neuromorphic bio-signal processing systems built following this approach adhere to design principles that are based on those of biological nervous systems~\cite{Sterling_Laughlin15}.
Their electronic circuits are typically designed using mixed-mode analog/digital transistors and fabricated using standard VLSI processes, to emulate the physics of real neurons and synapses in real-time~\cite{Chicca_etal14,Indiveri_Sandamirskaya19}.
Similar to the neural processes they model, neuromorphic systems process information using energy-efficient asynchronous, event-driven, methods.
They are adaptive, fault-tolerant, and can be flexibly configured to display complex behaviors by combining multiple instances of simpler elements.
The most striking difference between neuromorphic systems and conventional bio-signal information processing systems is in their unconventional (beyond von Neumann) computing architecture: rather than implementing one or more digital, time-multiplexed, central processing units physically separated from the main memory areas, they are characterized by co-localized memory and computation. Neuromorphic systems comprise bio-physically realistic neuron and synapse circuits that are at the same time the site of memory storage and of complex non-linear operations which perform collective and distributed computation.
This in-memory computing feature is the main reason that allows neuromorphic systems to perform bio-signal processing using orders of magnitude less power than conventional electronic computing systems~\cite{Indiveri_Liu15,Rahimiazghadi_etal20}.
Examples of ultra-low power brain-inspired neuromorphic systems able to process temporal patterns have been recently applied to \ac{EMG} signal processing~\cite{Ma_etal20,Donati_etal19} and \ac{ECG} anomaly detection~\cite{Bauer_etal19,Das_etal18a}.

Starting from the faithful emulation of individual neurons~\cite{Abu-Hassan_etal19}, neuromorphic processing systems can implement complex computations by assembling and combining multiple computational primitives, very much like standard computing based on Boolean logic is built from the combination of logic gates~\cite{Marcus_etal14,Chicca_etal14}. In particular, a computational primitive that is useful for generating a rich set of complex movements and switching behaviors is the \ac{CPG}. \acp{CPG} are biological neural circuits that produce rhythmic outputs to drive stereotyped motor behaviors like breathing, or chewing, walking, swimming and flying. Thanks to their simple structure and the limited numbers of parameters, neuromorphic implementations of \acp{CPG} have been applied to different fields as robotics~\cite{Donati_etal14, Gutierrez-Galan_etal20} and biomedical applications~\cite{Abu-Hassan_etal20}. 

In this paper we show how a neuromorphic implementation of neural oscillators is able to reproduce a sequence of population activation as shown in the biological \ac{rCPG} or in the heart chambers activation. An neuromorphic implementation of \ac{rCPG} can replace its biological counterpart in the vagus nerve stimulation to generate the desired heart rhythm. Otherwise, bypassing the biological \ac{rCPG}, the neural oscillator can directly generate stimuli for the heart chambers activation. In both cases, the network would use physiological input, such as the concentration of oxygen or carbon dioxide in the blood to modulate the heart pacing.

\section{METHODS}
\label{sec:methods}

\subsection{Mixed-signal neuromorphic processors}
\label{ssec:chips}

Unlike digital simulators of neural networks, mixed-signal neuromorphic processors use the physics of silicon to directly emulate neural and synaptic dynamics~\cite{Chicca_etal14}. In this case, the state variables evolve naturally over time and ``time represents itself''~\cite{Indiveri_Sandamirskaya19}, bypassing the need to have clocks and extra circuits to manage the representation of time.
Examples of neuromorphic architectures that follow this mixed-signal approach include the  Recurrent On-Line Learning Spiking (ROLLS) neuromorphic processor~\cite{Qiao_etal15}, and the Dynamic Neuromorphic Asynchronous Processor- scalable (DYNAP-SE) chip~\cite{Moradi_etal18}.

\begin{figure}
  \centering
  \includegraphics[width=\columnwidth]{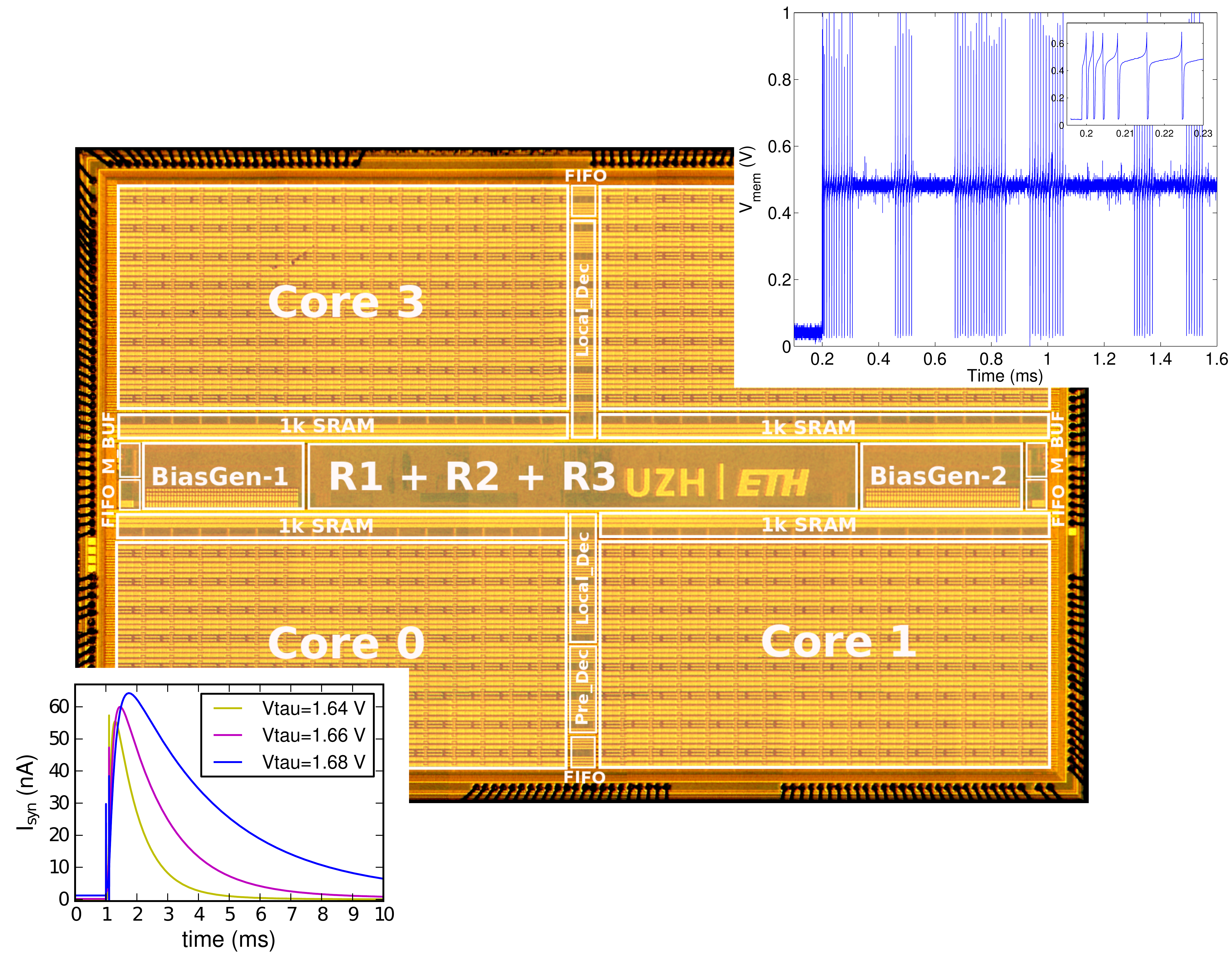}
  \caption{Neuromorphic processor micro graph with typical silicon synapse and silicon neuron response curves. The neuromorphic processor is the 4-core chip presented in~\cite{Moradi_etal18}. The lower left plot shows the impulse response of a single neuromorphic analog synapse circuit and the top right plot shows the response of a silicon neuron tuned to exhibit spike-frequency adaptation and bursting behavior. The synapse and neuron circuit details are presented in~\cite{Chicca_etal14}.} 
        \label{fig:dynapse}
\end{figure}

In these devices, the silicon neurons circuits implement a model of the adaptive exponential \ac{IF} neuron\cite{Brette_Gerstner05}, and the synapses implement biologically realistic first order synaptic dynamics~\cite{Chicca_etal14}. The circuits parameters can be tuned to make them exhibit linear and non-linear behaviors such as spike-frequency adaptation, refractory period saturation, regular firing, or bursting (see Fig.~\ref{fig:dynapse}).

\subsection{Oscillation generation}
\label{ssec:oscillation}
\begin{figure}[t]
   \centering
   \includegraphics[width=0.6\columnwidth]{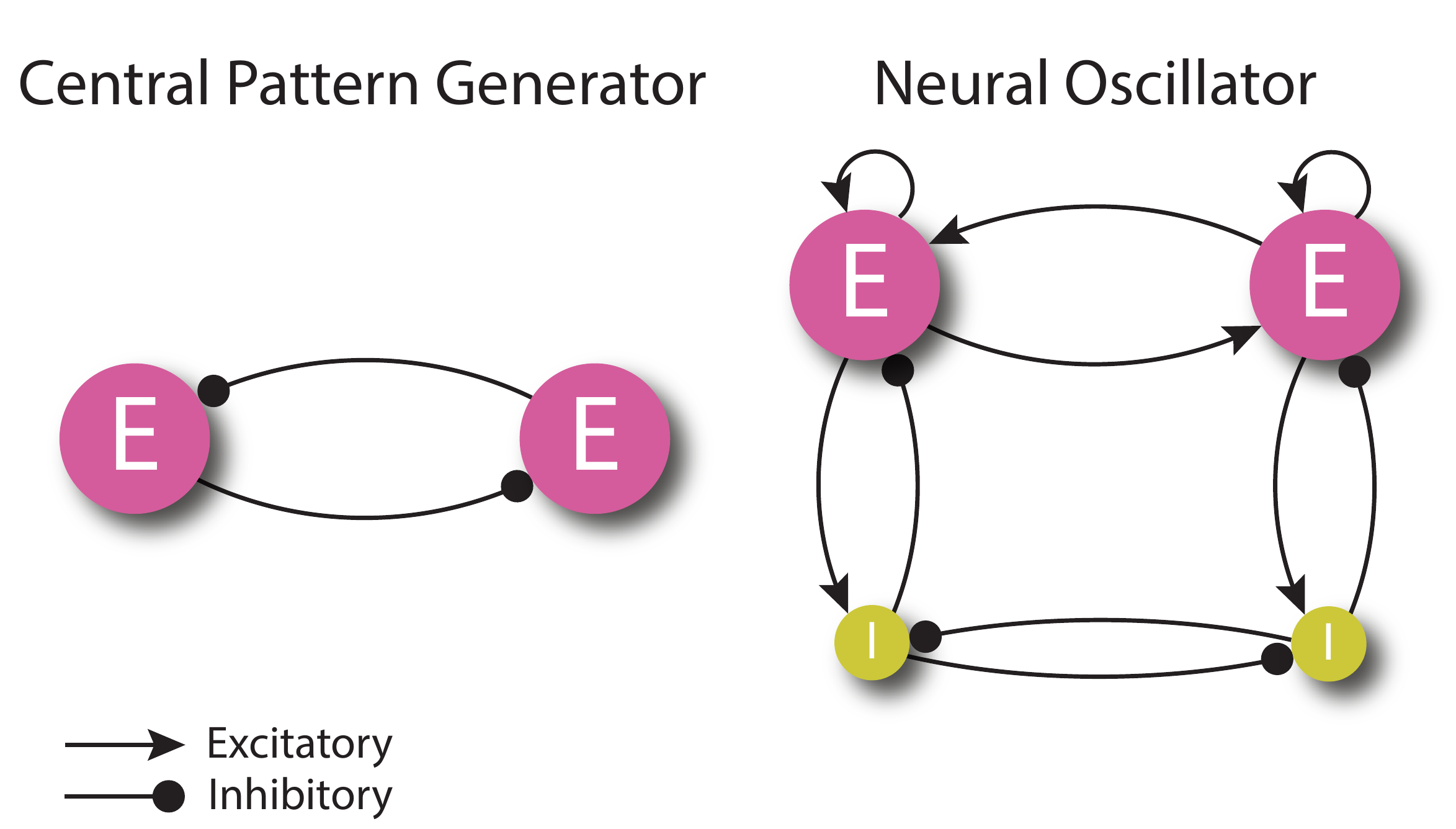}     \caption{Computational primitives to generate oscillations: the central pattern generator \ac{CPG} and the neural oscillator. The \ac{CPG} architecture (left) is composed of two excitatory populations coupled by mutual inhibition. The neural oscillator architecture (right) comprises two excitatory and two inhibitory populations mutually coupled. It follows Dale's law to be more biologically realistic and to provide a higher flexibility in tuning the oscillation properties.}
        \label{fig:architectures}
\end{figure}

A neural oscillation is a rhythmic or repetitive pattern of activity in the central nervous system. A basic building block for generating sequences with identical cycles such as those in the \ac{rCPG} or in the heart chambers activation is the ``neural oscillator''. In this section we present a neuromorphic implementation of a neural oscillator as an extension of a basic \ac{CPG} (see Fig.~\ref{fig:architectures}).

\subsubsection{Central Pattern Generation architecture}
\label{ssec:CPG}

\acp{CPG} are central nervous system networks that can generate coordinated muscle outputs in the absence of continuing patterned sensory input. Although \acp{CPG} are able to produce autonomously the desired rhythmic patterns, their activation is constantly monitored and modulated by high level centers. Together with the constant sensory feedback, modulatory top-down inputs allow high adaptability to proprioceptive and environmental conditions. As a result, the \ac{CPG} output is typically a spatio-temporal sequence pattern with phase lags between the temporal sequences that corresponds to different rhythms. 

Traditionally, this concept was applied to simple, innate, rhythmic movements with identical cycles that repeat continually (e.g. respiration) or irregularly (e.g. locomotion). More recent studies showed that many natural movement sequences are not simple rhythms, but include different elements in a complex order where some involve learning~\cite{Berkowitz19}.

A key model to understanding rhythm generation is the half-center oscillator network. It consists of two neurons that have no rhythmogenic ability individually, but produce rhythmic outputs when reciprocally coupled. In a \ac{CPG} two pools of neurons are mutually coupled with inhibitory connections. Neurons of each pool make transition between activated and inhibited phases following an ``escape'' or a ``release'' mode. In the release mode the inhibited neurons can escape from inhibition thanks to the intrinsic properties of the membrane and inhibit the other neurons whereas in the escape one the neurons show spike-frequency adaptation and slowly stop firing, releasing the other inhibited neurons. 

\subsubsection{Neural oscillator architecture}
\label{ssec:cosc}
An alternative \acp{CPG} architecture is that of a neural oscillator, which can produce rhythmic outputs without resorting to adaptation features of the neurons.
Here we refer to a neural oscillator consisting of an excitatory and a reciprocally connected inhibitory neuron population.
The neurons of the excitatory population are connected to each other leading to self-excitatory behavior of the excitatory population and receive a constant input which drives the neural oscillator.

The neural oscillator architecture leads to limit cycle behavior due to the reciprocal connections between the excitatory and the inhibitory neuron populations.
The constant input to the excitatory neuron population leads to an increase in the activity of the excitatory population.
The excitatory population then starts to activate its inhibitory counterpart which in return will inhibit the excitatory population and reset the network.
Similar to the \acp{CPG} architecture with adaptation, here the strength of the inhibition is proportional to the activity level of the excitatory neurons and thereby leads to these oscillatory dynamics.

Neural oscillators are a useful building block for systems which generate rhythmic outputs.
They can be coupled by adding connections between their excitatory populations and another set of connections between their inhibitory populations.
This coupling of individual neural oscillators allows it to precisely tune the phase shift between the individual neural oscillators.
Hence, by coupling multiple neural oscillators among each other and tuning their phase shifts accordingly, it is possible to obtain a large variety of complex, periodic output patterns.

\begin{figure}
  \centering
  \includegraphics[width=0.55\columnwidth]{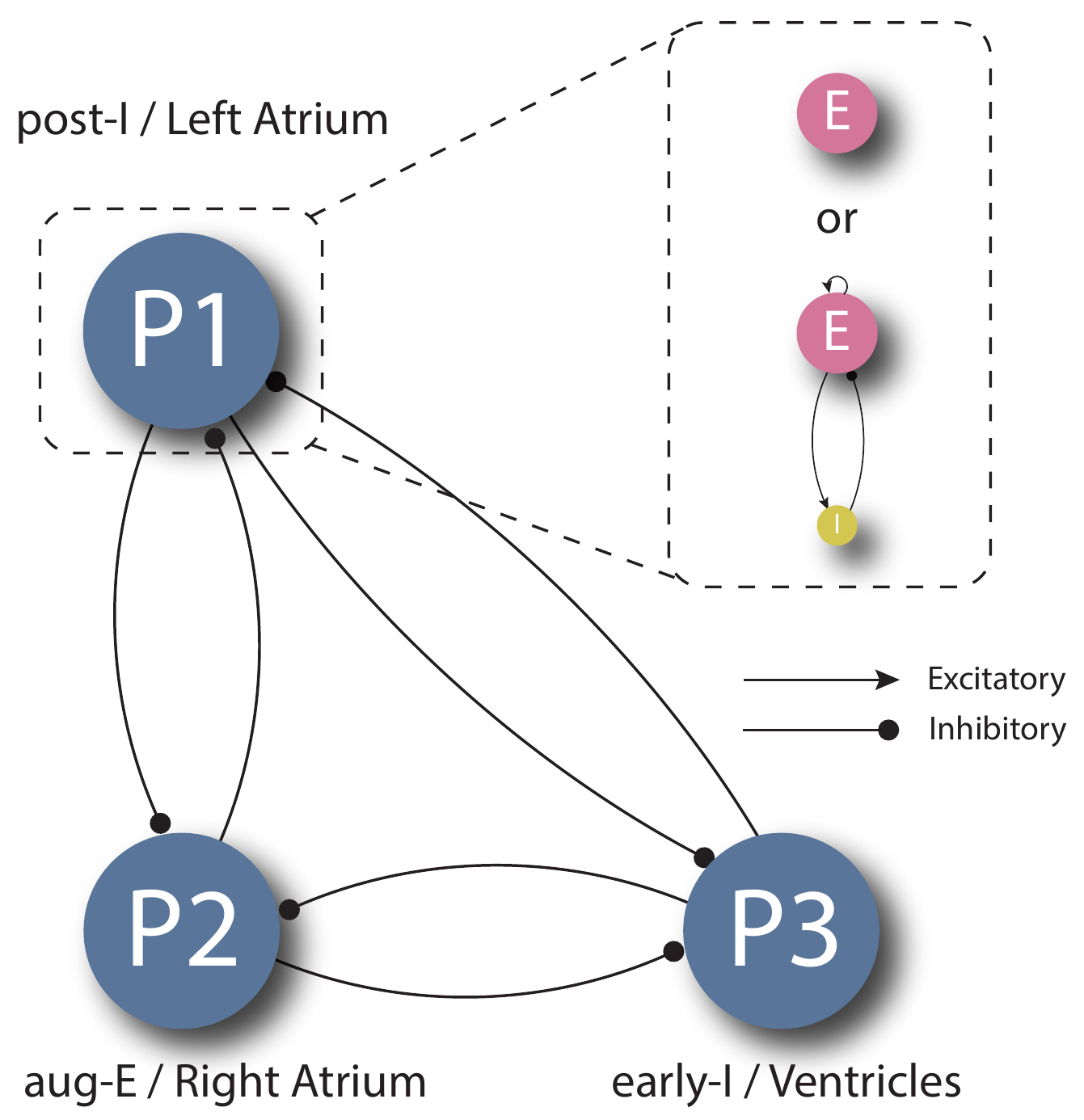}  
  \caption{Three-phase network, to model the \ac{rCPG} as well as the heart chambers activation. Each population can be modelled using two different architectures: the \ac{CPG} unit and the neural oscillator.}
  \label{fig:resp_net}
\end{figure}

\subsection{Three-phase network}
\label{ssec:3phase_net}
We model the rCPG as well as the heart chambers activation as a three-phase network.
For the rCPG the three-phase network directly represents its biological counterpart in the medulla oblongata.
The medullary network operates in a three-phase rhythm which consists of the early inspiration phase (early-I), the post inspiration phase (post-I) and the late inspiration phase (aug-E)~\cite{Smith_etal07}.
We also model the heart chambers activation as a three-phase network because it is general practice to only stimulate a subset of the heart chambers.
Our cardiac three-phase network model does not aim to reproduce the mechanism of the heart but only to reproduce the rhythmic output.
It generates the stimulation time of the two atria separately and ventricles combined.
This allows us to show the hardware's ability to generate very short (25-30\,ms) as well as relatively long (800-900\,ms) delays between different activation phases which are the upper and lower boundaries of delays required for any other heart stimulation pattern.

\section{RESULTS}
\label{sec:results}

\subsection{Central Pattern Generation results}
\label{ssec:cpg_results}
\begin{figure}
  \centering
  \includegraphics[width=0.65\columnwidth]{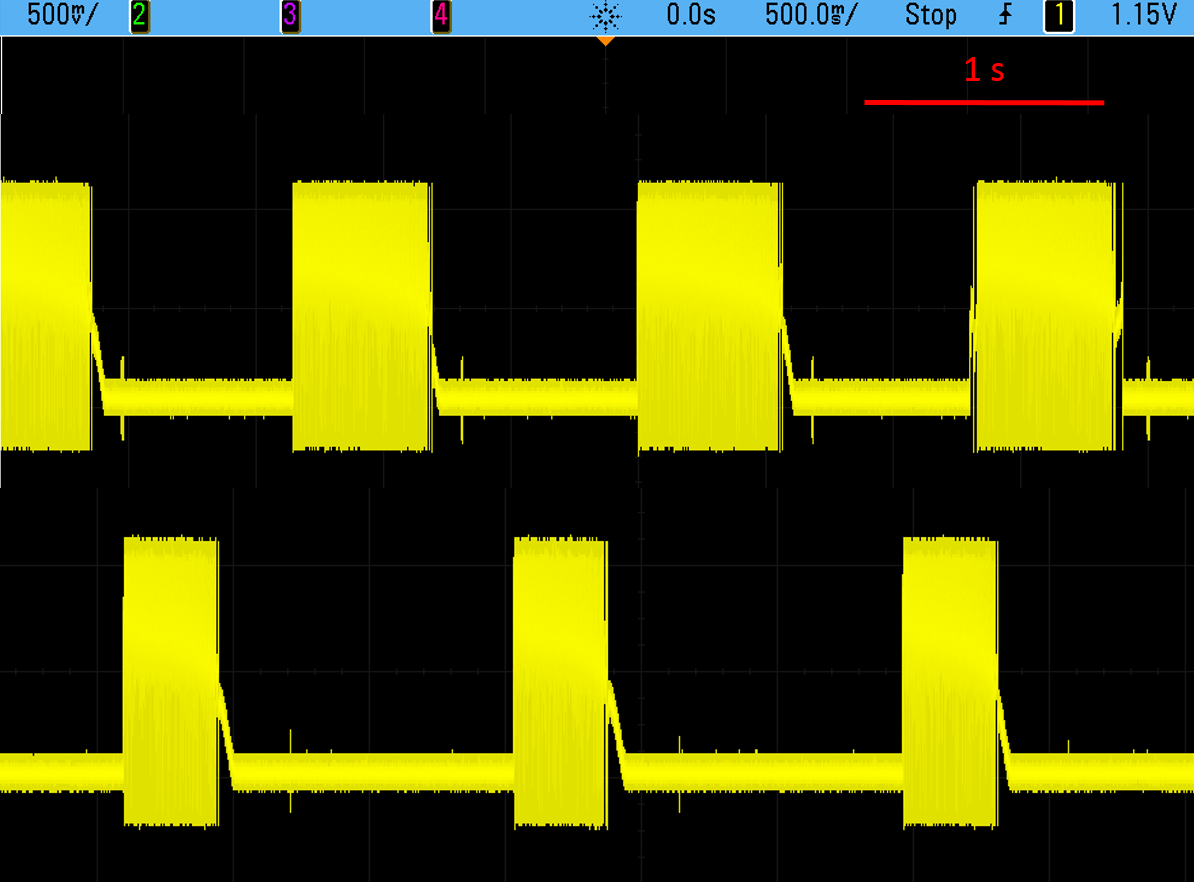}  
  \caption{Silicon neuron measurements. The neuron produces periodic bursting, lasting approximately 500 ms, with an equally long  inter-burst interval.}
  \label{fig:burst}
\end{figure}

The first step to reproduce the \ac{CPG} behavior is to set the correct parameters of neuron and synapse circuits to generate bursting dynamics. This burst activity can be reproduced by setting strong adaptation current in silicon neurons and weak synaptic weights and slow time constants in the synapses, Fig.~\ref{fig:burst}. 

\begin{figure}
  \centering
  \includegraphics[width=0.85\columnwidth]{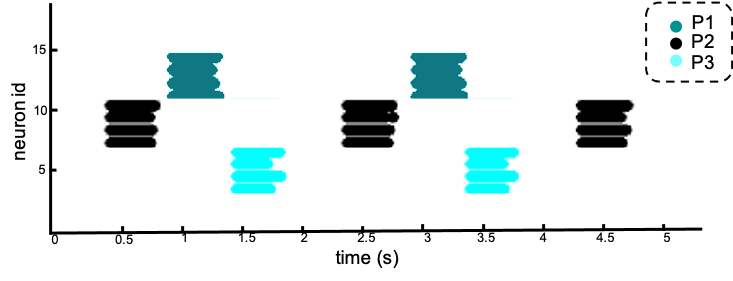}  
  \caption{Raster plot of the three-phase network with \ac{CPG}. The network consists of 3 excitatory populations (n=4) and oscillates with a frequency of 1\,Hz}
  \label{fig:3cpg}
\end{figure}

In the neuromorphic \ac{CPG} the frequency obtained for a single unit alternate movements ranges from $0.5 Hz$ to around $3.0\,Hz$. 
Figure~\ref{fig:3cpg} shows the raster plot of three segments, P1, P2, P3, at low frequency, around $0.5\,Hz$. This frequency allows the emulation of the basic \ac{rCPG}, as well as, heart chambers activation rhythms. 

\subsection{Neural Oscillator results}
\label{ssec:no_results}
\begin{figure}
  \includegraphics[width=\columnwidth]{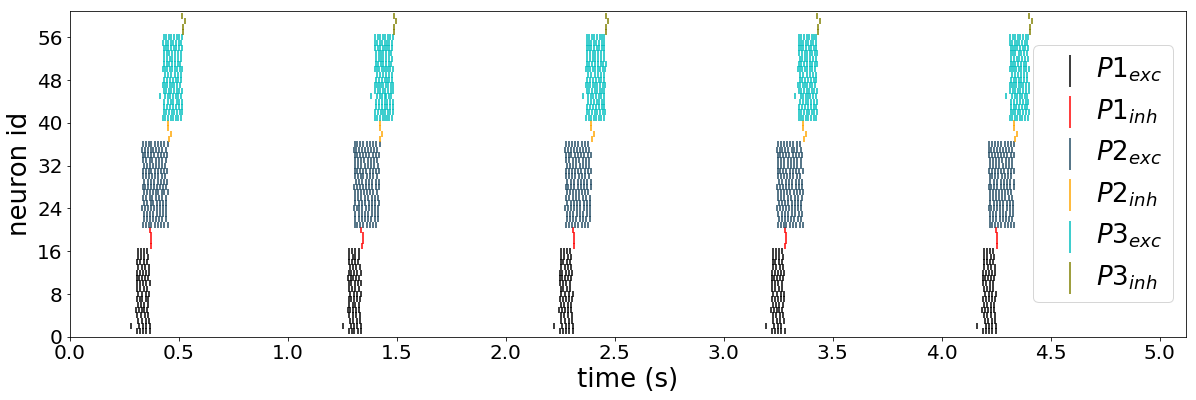}  
  \caption{Raster plot of the three-phase network with neural oscillators. The network consists of 6 neuron populations (3 excitatory (n=16) and 3 inhibitory populations (n=4)) and oscillates with a frequency of $1\,Hz$.}
  \label{fig:cosc_res}
\end{figure}

The oscillation frequency of a neural oscillator is directly proportional to the constant input current to the excitatory population.
Here we tuned our system of three coupled neural oscillators to run at a frequency of roughly $1\,Hz$ (see Fig.~\ref{fig:cosc_res}) on the DYNAP-SE chip~\cite{Moradi_etal18}.

\begin{figure}[h]
  \includegraphics[width=\columnwidth]{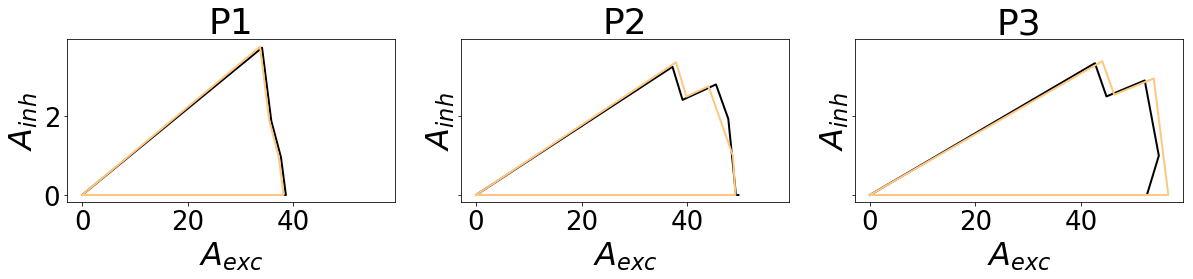}
  \caption{Phase diagram of neural oscillators. 
The axes indicate the average activity of the excitatory (x-axis) resp. inhibitory (y-axis) population of the three neural oscillators over two oscillation cycles (Cycle1: black, Cycle2: yellow).}
  \label{fig:phase}
\end{figure}

The noise and device mismatch in mixed-signal analog/digital neuromorphic circuits leads to small variations in the population activities of the neural oscillators over time.
This is illustrated in Fig.~\ref{fig:phase} which shows the phase diagram of each neural oscillator in the three-phase network for two adjacent oscillation cycles.
All three neural oscillators are similarly affected by the noise which is shown by the similar deviations between the yellow and black lines for P1, P2 and P3 while they remain fully synchronized as shown in Fig,~\ref{fig:cosc_res}.
Overall, the three-phase network has a standard deviation of roughly $2\,ms$ for an oscillation frequency of $1\,Hz$.

\section{CONCLUSIONS}

\label{sec:conclusions}
In  this  paper  we  introduced the neuromorphic engineering approach applied to bioelectronic medicine and presented a neuromorphic implementation of a three-phase network that emulates a \ac{rCPG} and/or a heart chambers activation sequences, as a key technology for adaptive pacemakers. The cycle in the three-phase can be generated by using two different computational primitives, the \ac{CPG} unit or the neural oscillator. We show how both primitives can be used to exhibit similar results, with oscillation frequencies of approximately 1\,Hz, compatible with the frequencies involved in the \ac{rCPG} and cardiac rhythm.

\addtolength{\textheight}{-12cm}  
\bibliographystyle{IEEEtran}
\bibliography{biblioncs}

\end{document}